%% file: main.tex
\documentclass{article}

\usepackage[utf8]{inputenc}
\usepackage[T1]{fontenc}

\PassOptionsToPackage{table, dvipsnames}{xcolor}
\usepackage{xcolor}

\PassOptionsToPackage{colorlinks=true, citecolor=blue, linkcolor=blue, urlcolor=black}{hyperref}

\usepackage{arxiv}
\usepackage[utf8]{inputenc}
\usepackage[T1]{fontenc}
\usepackage{url}
\usepackage{booktabs}
\usepackage{nicefrac}
\usepackage{microtype}
\usepackage{lipsum}
\usepackage{graphicx}
\usepackage{natbib}
\usepackage{doi}
\usepackage{algorithm}
\usepackage{multirow}
\usepackage{multicol}
\usepackage{amsmath,amssymb,amsfonts}
\usepackage{amsthm}
\usepackage{mathrsfs}
\usepackage{float}
\usepackage{algorithmic}
\usepackage{bm}
\usepackage{caption}
\usepackage{hyperref}

\title{High-resolution velocity model estimation with neural operator and the time-shift imaging condition}

\author{ \href{https://orcid.org/0009-0009-0709-9158}{\includegraphics[scale=0.06]{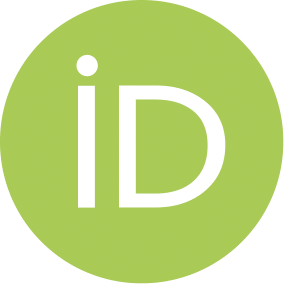}\hspace{1mm}Xiao~Ma}\\
	Division of Physical Science and Engineering\\
	King Abdullah University of Science and Technology\\
	Thuwal 23955-6900, Saudi Arabia \\
	\texttt{xiao.ma@kaust.edu.sa} \\
        \And
	\href{https://orcid.org/0000-0001-7027-5956}{\includegraphics[scale=0.06]{orcid.png}\hspace{1mm}Shaowen~Wang} \\
	Division of Physical Science and Engineering\\
	King Abdullah University of Science and Technology\\
	Thuwal 23955-6900, Saudi Arabia \\
	\texttt{shaowen.wang@kaust.edu.sa} \\
            \And
	\href{https://orcid.org/0000-0002-9363-9799}{\includegraphics[scale=0.06]{orcid.png}\hspace{1mm}Tariq~Alkhalifah} \\
	Division of Physical Science and Engineering\\
	King Abdullah University of Science and Technology\\
	Thuwal 23955-6900, Saudi Arabia \\
	\texttt{tariq.alkhalifah@kaust.edu.sa} \\
}

\renewcommand{\shorttitle}{Time-lag RTM enhanced VMB using Neural Operator}
\graphicspath{{./Figures/}}
\hypersetup{
pdftitle={A template for the arxiv style},
pdfsubject={q-bio.NC, q-bio.QM},
pdfauthor={David S.~Hippocampus, Elias D.~Striatum},
pdfkeywords={First keyword, Second keyword, More},
}

\begin{document}
\maketitle

\input{Sections/Abstract}
\input{Sections/Introduction}

\input{Sections/Method}
\input{Sections/Numerical_Examples}

\input{Sections/Conclusions}

\input{Sections/Acknowledgment}
\input{Sections/CodeAvailability}

\bibliographystyle{unsrtnat}
\bibliography{references}

\end{document}

%% file: Sections/Abstract.tex
\begin{abstract}
Extracting subsurface velocity information from seismic data is mainly an undetermined problem that requires injecting a priori information to constrain the inversion process. Machine learning has offered a platform to do so through the training process, as we formulate our training dataset to inject as much prior knowledge as possible in the trained ML model. Here, we use a neural-operator-based framework for high-resolution seismic velocity model building, which integrates Neural Operators with time-lag reverse time migration imaging. Unlike conventional full waveform inversion (FWI) methods that rely on iterative forward and adjoint-state computations, our approach learns a direct mapping from initial velocity models and extended seismic images to high-resolution velocity estimates through supervised learning. The network architecture enables mesh-independent generalization by learning mappings in infinite-dimensional function spaces, while synthetic velocity models and time-lag reverse time migration (RTM) images provide complementary high-frequency information critical for recovering mid- and high-wavenumber velocity components. Synthetic experiments demonstrate that the proposed method accurately reconstructs fine-scale structures and complex geologies, including out-of-distribution features such as salt bodies. Applications to real seismic field data acquired offshore Australia further validate the method’s robustness and resolution capability. The predicted models show enhanced structural details and improved consistency with well-log data, outperforming traditional multi-scale FWI in both accuracy and computational efficiency. The entire prediction process is completed within seconds, making the proposed approach highly suitable for rapid and scalable velocity model building in practical exploration scenarios.
\end{abstract}

\keywords{Deep learning \and Neural operators \and Reverse time migration \and Automatic differentiation}

%% file: Sections/Introduction.tex
\section{\textbf{Introduction}}
In recent years, there has been a growing interest in the application of deep learning-based algorithms for seismic velocity model building, driven by the increasing availability of large-scale seismic datasets and computational resources. \textcolor{black}{These deep learning-based algorithms, unlike traditional full waveform inversion (FWI) methods, which require the use of the adjoint-state method to compute gradients, offer near instant predictions, as they aim to bypass traditional iterative inversion procedures} by learning direct mappings from seismic data to subsurface velocity structures. \cite{araya2018deep} first demonstrated that convolutional neural networks (CNNs) and fully convolutional architectures can be trained to convert multi-shot seismic data into detailed velocity structure, offering a significant reduction in computational cost compared to conventional methods. In addition, \cite{yang2019deep} extended this approach to the task of salt body inversion and observed that, compared to conventional FWI, the neural network–based method yields a more accurate delineation of the salt body. \cite{kazei2020velocity} applied the direct inversion approach to field seismic data, demonstrating its practical feasibility and potential for real-world applications. Another work seeks to generate background velocity models suitable for use in Full Waveform Inversion (FWI). These models are typically smoother and aim to capture low-wavenumber background velocity that guides the subsequent inversion process \citep{farris2018tomography,mohamed2023use}. \par

The aforementioned methods primarily rely on CNNs, which rely on local convolutional operators. However, in geophysical scientific computing, training data are typically generated using conventional numerical solvers \citep{berryhill1979wave,mora1987nonlinear}. Producing high-quality datasets through these solvers can be extremely time-consuming, often requiring days or even months of computation \citep{harsuko2025propagating}. In some cases, generating sufficiently diverse and accurate training samples is not feasible. Moreover, neural networks are often viewed as interpolators with limited extrapolation capabilities, raising concerns about their ability to generalize to unseen scenarios. For instance, if the training data are generated at a fixed resolution, the learned models may fail to generalize to problems defined on different discretizations. Overall, generalization remains a fundamental challenge in geophysical machine learning. This introduces a trade-off: while machine learning-based methods can significantly accelerate evaluation or inference, their training phase can be computationally intensive and less flexible. To solve these problems, several recent studies have proposed the use of Transformer-based architectures \citep{wang2023seismic} and neural operators \citep{lu2021learning,molinaro2023neural} to improve the accuracy and effectiveness of inversion results. In particular, neural operators, as an emerging class of network architectures, have shown remarkable success in supervised learning tasks \citep{li2020neural,kovachki2023neural, cao2024laplace}. Unlike conventional neural networks that approximate mappings between finite-dimensional vector spaces, neural operators are specifically designed to learn mappings between infinite-dimensional function spaces. This formulation allows them to model complex physical systems, such as partial differential equations (PDEs), in a mesh-independent and data-efficient manner. Among them, the Fourier Neural Operator (FNO) stands out due to its use of global convolution via the Fast Fourier Transform (FFT), enabling it to effectively capture long-range dependencies and non-local interactions \citep{li2020fourier}. For example, \cite{yang2021seismic} employed the FNO to learn the two-dimensional acoustic wave equation. Experimental results demonstrate that even with a limited amount of training data, the FNO is capable of producing forward modeling results comparable to those obtained using traditional numerical methods, across varying velocity models and source locations. Furthermore, \cite{wei2022small} extended the approach to elastic wave propagation, demonstrating its potential in more complex physical systems. In addition to solving forward modeling problems, some studies \citep{zou2024deep, zou2025ambient} have also leveraged the differentiability of neural operators to perform FWI through automatic differentiation. However, the application of such methods to field data remains challenging and is still subject to several practical limitations. Notably, \cite{huang2023automated} proposed the use of an FNO to directly predict initial velocity models for FWI from raw shot gathers. Although the model is trained solely on synthetic seismic data, tests conducted on field data demonstrate that the FNO is still capable of producing reasonably accurate background velocity models. Based on this work, \cite{crawley2023high,crawley2024shortening,korsmo2025velocity} proposed feeding both the background velocity model and angle gathers into FNO. In this framework, the angle gathers serve as a source of high-wavenumber information, complementing the low-frequency content of the background velocity. Through multiple field-data experiments, the results demonstrate that the neural operator effectively integrates \textcolor{black}{such information} to produce high-resolution velocity models, highlighting its potential for improved subsurface imaging in complex geological settings. \par

A challenge in applying neural networks to geophysical problems is ensuring their ability to generalize effectively from synthetic training data to real-world field data. Neural networks often exhibit performance degradation when the features of test data deviate from those during training. Several approaches have been proposed to address the gap between synthetic training data and field data, including style transfer \citep{takemoto2019enriching, ovcharenko2019style,el2022styletime}, domain adaptation \citep{alkhalifah2021mlreal,birnie2022leveraging}, and data augmentation \citep{wang2021seismogen,bonke2024data}. In seismic migration, the use of extended images can also be regarded as a form of data augmentation. By incorporating time-lag or space-lag dimensions, extended images implicitly encode additional information about subsurface reflectivity \citep{bednar2005brief}. This enriched representation enhances the diversity and expressiveness of the training data. The neural operator is capable of learning from these extended features, effectively extracting hidden structural content that may not be directly accessible from conventional inputs. Moreover, by analyzing and extracting key structural patterns—such as layer geometries, fault orientations, or salt body shapes—from migrated field images or background models, we can generate synthetic velocity models that mimic the geological features observed in practice. To some extent, this strategy can be viewed as a form of structure-aware domain adaptation, where the style or structural characteristics of field data guide the generation of more realistic and domain-relevant synthetic training examples. \par

In this paper, first, we provide an overview of the adapted Neural Operator architecture, along with the workflow used to generate the synthetic samples for training. Then, we test the performance of our neural operator using in-distribution and out-of-distribution data. Subsequently, we present a successful application of the trained model to a field dataset acquired in Australia.

%% file: Sections/Method.tex
\section{Theory}
In this section, we first introduce the concept of operator learning. We then describe the workflow for constructing the synthetic training dataset, including the design of velocity models and the generation of corresponding time-lag seismic images. Finally, we present the architecture of the proposed neural operator, highlighting its key components and modifications suitable for the seismic inversion task.
\subsection{Operator learning theory}

Neural operators (NOs) \citep{li2020neural} represent a novel class of neural network architectures developed to learn mappings between infinite-dimensional function spaces. Unlike conventional neural networks, which are typically constructed to approximate mappings between high-dimensional but finite vector spaces, neural operators are explicitly designed to operate within the functional space framework. To this end, NOs incorporate architectural mechanisms that ensure the learned mappings are robust with respect to variations in the discretization of the input functions. 

In the context of learning solution operators for families \textcolor{black}{integral or} Partial Differential Equations (PDEs), neural operators aim to learn the mapping
\begin{equation}
\mathcal{G}_\theta: \mathcal{A} \rightarrow \mathcal{U},
\end{equation}
where $\theta \in \Theta$ represents the parameterization of \textcolor{black}{the solution operator (the solver)}, $\mathcal{A}$ is a function space representing the initial or input conditions, and $\mathcal{U}$ is a function space containing the solution function space~\citep{li2020fourier,li2020neural}.

To approximate the target operator $\mathcal{G}$, neural operators adopt an iterative architecture that constructs a sequence of intermediate functions $\bm{z}_1, \ldots, \bm{z}_T$ within a lifted representation space. In this context, "lifting" refers to the transformation of the input function into an initial feature map $\bm{z}_1: \mathbb{R}^{d} \rightarrow \mathbb{R}^{c}$ via a pointwise operation, commonly implemented as an affine mapping followed by a nonlinear activation, applied independently at each spatial location $\bm{r} \in \mathbb{R}^d$. The final representation $\bm{z}_T$ is then projected back to the physical space as the network output $\mathcal{U}$, also through a pointwise transformation. The network is trained to ensure that the resulting function $\mathcal{U}$ accurately approximates the solution of the underlying problem in a suitable function norm, such as the $L^2$ norm.

The transition between successive representations is governed by the recursive update:
\begin{equation}
\bm{z}_{t+1}(\bm{r}) = \sigma \left( \bm{W}_t \bm{z}_t(\bm{r}) + \left( \mathcal{A}_{\psi} \bm{z}_t \right)(\bm{r}) \right),
\end{equation}
where $\sigma$ denotes a nonlinear activation function (e.g., ReLU or GELU), $\bm{W}_t$ is a trainable linear transformation acting along the channel dimension, and $\mathcal{A}_{\psi}$ represents a parameterized kernel integral operator with learnable parameters $\psi$.

In the case of Fourier Neural Operators (FNOs) used in this paper, the integral operator $\mathcal{A}_{\psi}$ is implemented using a spectral convolution defined by:
\begin{equation}
\left( \mathcal{A}_{\psi} \bm{z}_t \right)(\bm{r}) = \mathcal{F}^{-1} \left( \mathcal{M}_{\psi} \cdot \left( \mathcal{F} \bm{z}_t \right) \right)(\bm{r}),
\end{equation}
where $\mathcal{F}$ and $\mathcal{F}^{-1}$ denote the Fourier and inverse Fourier transforms, respectively, and $\mathcal{M}_{\psi}$ is a learnable modulation mask acting in the frequency domain. This mask includes a low-pass filter and a channel-wise multiplicative operator parameterized by $\psi$ \citep{li2020fourier}. The number of retained Fourier modes per dimension in $\mathcal{M}_{\psi}$ is controlled by a user-defined hyperparameter, thereby regulating the resolution and expressivity of the spectral representation.

\subsection{Synthetic velocity and corresponding time-lag migration images}

\begin{figure}[H]
\centering
\includegraphics[width=\textwidth]{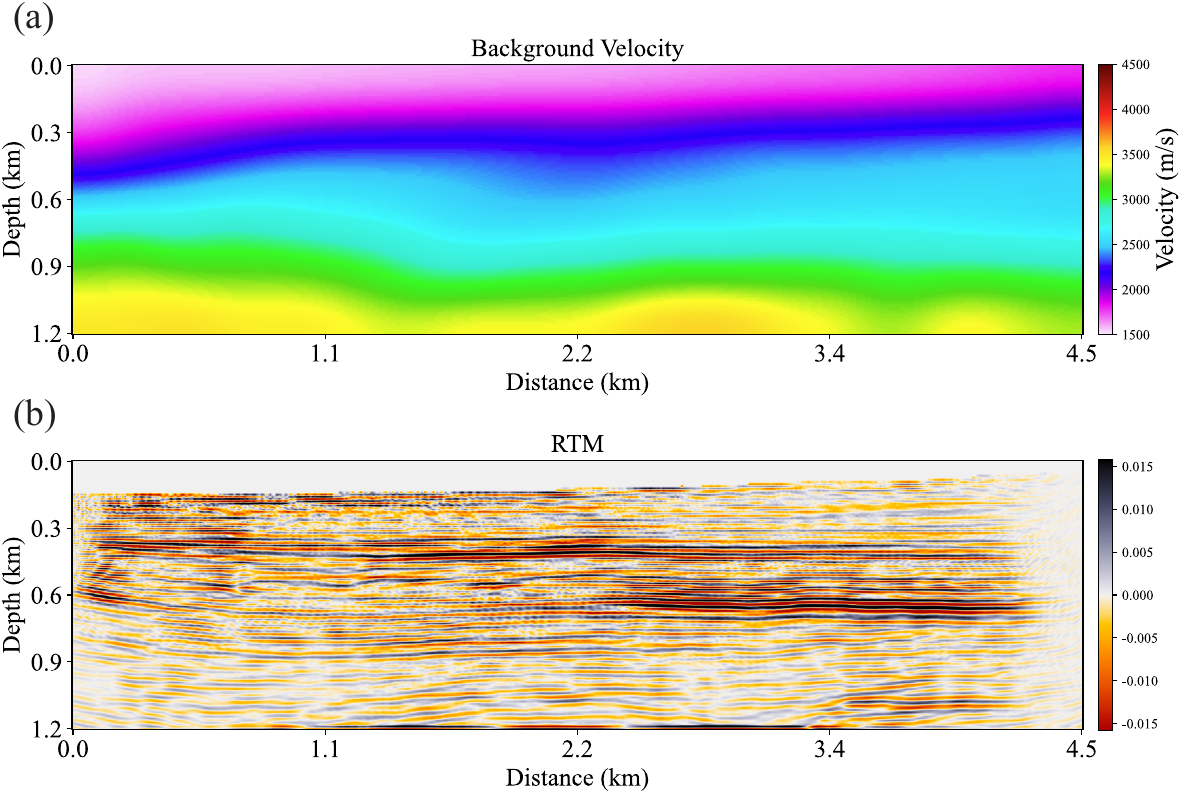}
\caption{(a) and (b) show the background velocity model and the corresponding RTM image derived from field data, respectively. The structural features evident in the RTM image serve as valuable references for constructing synthetic seismic datasets used in training.}
\label{fig1} 
\end{figure}

In this study, we employ a supervised learning framework to train the neural operator. A key aspect of our methodology is the construction of synthetic velocity models that reflect the structural characteristics observed in the field data. Specifically, as shown in Figure~\ref{fig1}, the RTM image of the field data exhibits relatively flat and laterally continuous reflectors, without prominent geological anomalies such as salt bodies or complex fault systems. Based on this observation, we construct synthetic velocity \textcolor{black}{samples} that preserve such structural simplicity. This approach provides a new perspective for generating synthetic models: instead of designing them arbitrarily, we first apply migrate field data to identify dominant structural features, and then use these features to guide the construction of synthetic velocity models. This target-informed strategy ensures that the synthetic models are geologically consistent with the target field data, thereby enhancing the relevance and effectiveness of supervised training for the neural operator. \par

\begin{figure}[H]
\centering
\includegraphics[width=\textwidth]{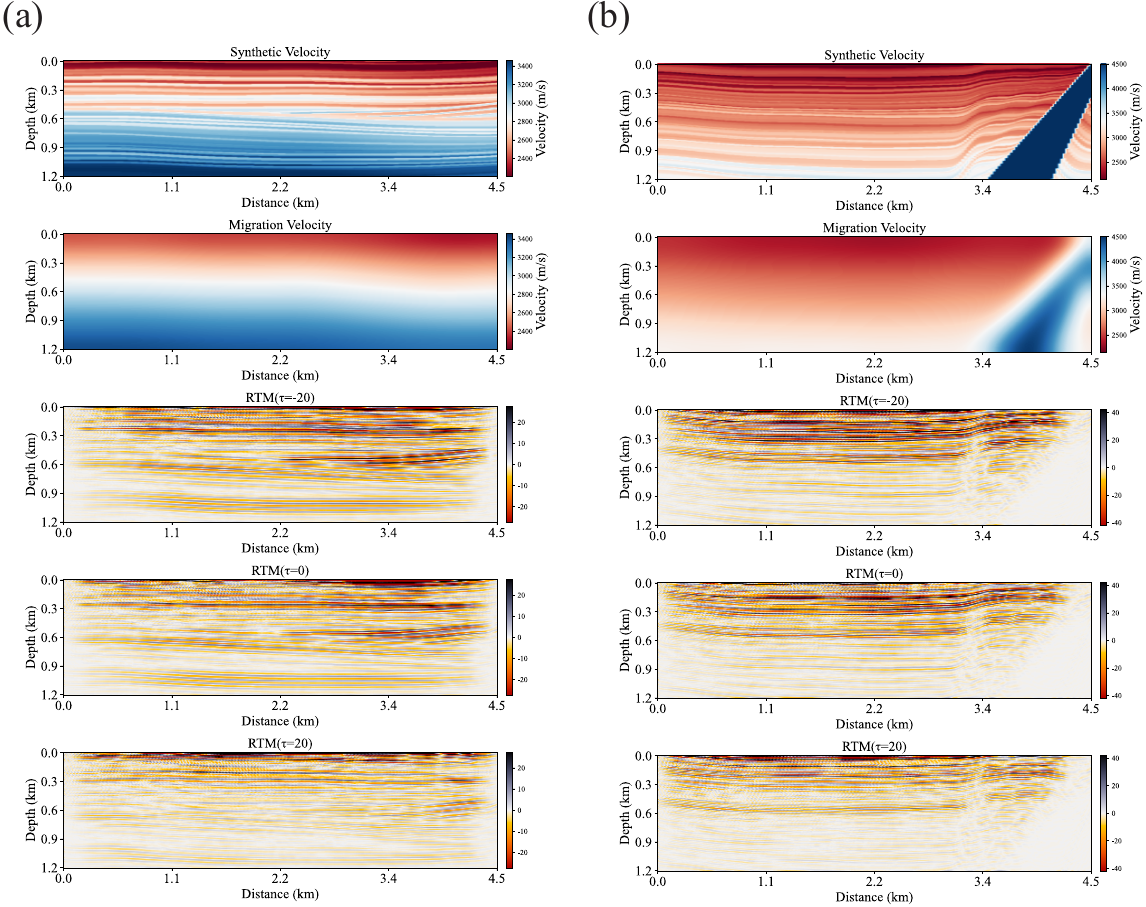}
\caption{\textcolor{black}{(a) and (b) illustrate two distinct types of synthetic velocity models along with their corresponding migration velocity and time-lag RTM images, samples from the training set. The migration velocity and the three RTM images form a four-channel input to the neural operator, while the true velocity acts as the label.}}
\label{fig2} 
\end{figure}

Moreover, we further employ the time-lag migration approach to construct multiple extended RTM images for each synthetic seismic dataset. This technique not only enriches the training samples by introducing additional dimensions (channels), but also improves the robustness of the resulting images to velocity model inaccuracies \citep{yang2011wave}. Specifically, the extended images are computed using the following correlation-based formulation \citep{sava2006time}:
\begin{equation}
I(\mathbf{x}, \tau) = \sum u_f(\mathbf{x}, t + \tau) \cdot u_b(\mathbf{x}, t - \tau),
\end{equation}
where $u_f$ and $u_b$ denote the forward- and backward-propagated wavefields, respectively, and $\tau$ represents the time lag. By incorporating such time-lag gathers, this method effectively serves as a proxy for multi-angle imaging that helps the neural operator capture subsurface reflectivity across a range of propagation paths. In Figure~\ref{fig2}, we present two synthetic velocity samples along with \textcolor{black}{the migration velocity and} their corresponding time-lag RTM images. \textcolor{black}{The migration velocity is obtained from the synthetic velocity model by applying Gaussian smoothing with a sigma of 15.} The structural design of both synthetic models is informed by key geological features extracted from real seismic RTM images (Figure~\ref{fig1}(b)). \textcolor{black}{To enhance the generalization ability of the neural network with respect to diverse geological settings, we further incorporate complex structures such as salt bodies into the models}. For each synthetic velocity model, we compute three RTM images using different time-lag values from the resulting synthetic data. \textcolor{black}{The RTM images are computed by using the true velcoity model sample to first forward model with a wavelet extracted from the field seismic data to simulate shot gathers. Subsequently, the corresponding migration velocity is also employed to obtain the corresponding RTM images. The acquisition geometry used in the forward modeling is consistent with that of the field data, which will be shared later in the field data section.} As illustrated, each time-lag image captures distinct aspects of the subsurface reflectivity, providing complementary information for the training stage. 

\subsection{Neural operator architecture}

\begin{figure}[H]
\centering
\includegraphics[width=0.9\textwidth]{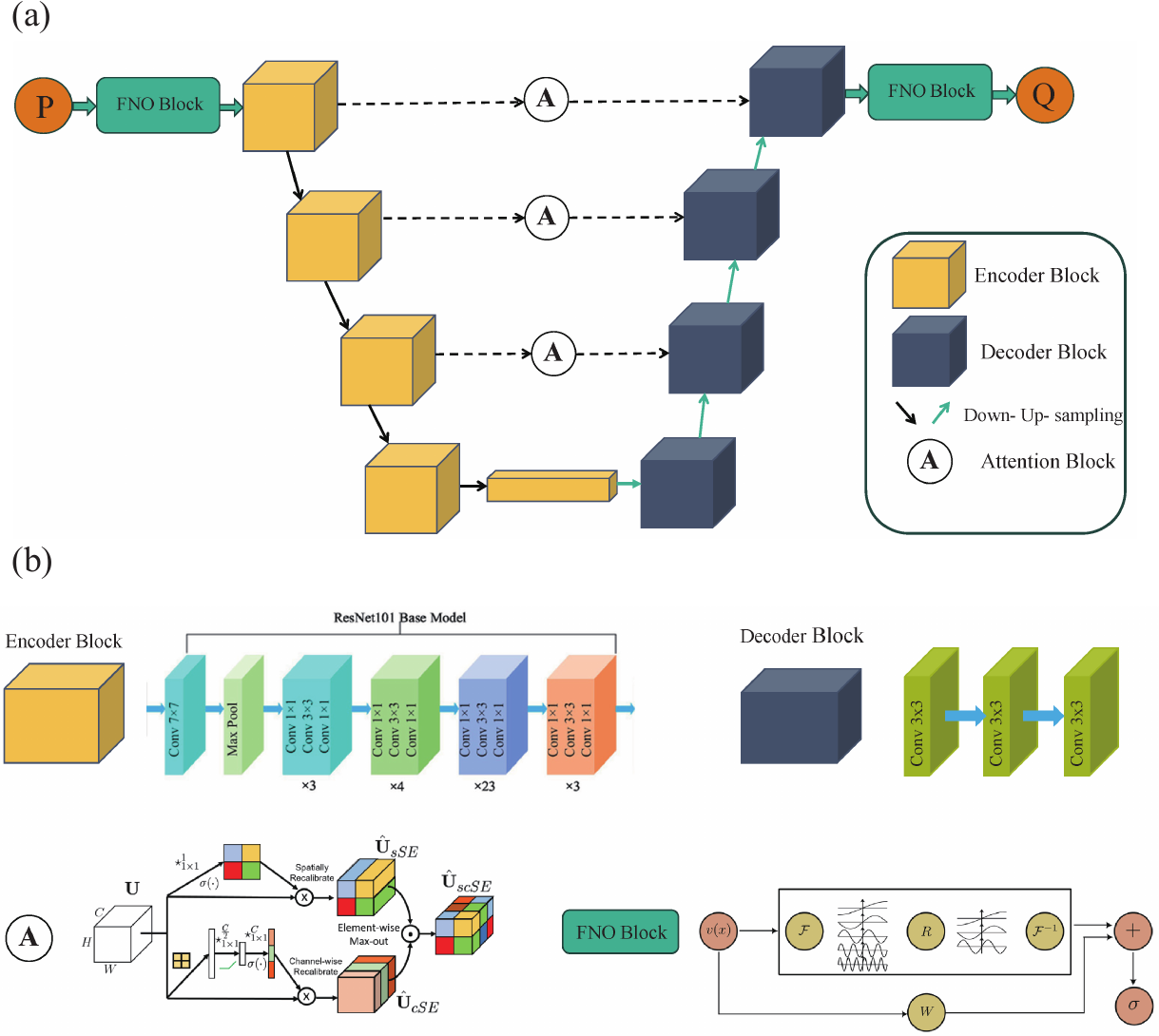}
\caption{(a) presents the overall architecture of the neural operator proposed in this study, including the Fourier operator layers and the encoder-decoder framework. (b) provides a detailed view of the internal structure of each individual component within the neural operator.}
\label{fig3} 
\end{figure}

The neural operator employed in this study is composed of Fourier operator layers and convolutional neural networks. The Fourier operator layer consists of a forward Fourier transform followed by an inverse Fourier transform. This transformation facilitates the modeling of global interactions and enhances the network’s ability to capture long-range dependencies, which are often challenging for purely spatial-domain methods. Specifically, the training data is first processed by a lifting function ($P$) followed by an FNO block, which together project the input data into a latent space. Subsequently, the FNO block is connected to an encoder-decoder style convolutional neural network architecture. Each encoder block is constructed based on the ResNet-101 \citep{he2016deep} baseline model, which provides a deep residual learning framework capable of capturing multi-scale hierarchical features. On the decoder side, each block consists of three convolutional layers with kernel sizes of 3×3. To further enhance the flow of information across the network, we incorporate both channel attention and spatial attention mechanisms within the skip connections \citep{mou2019cs}. These attention modules help the network focus on the most informative features by adaptively reweighting feature responses along both channel and spatial dimensions, thus improving the representational quality and robustness of the model.
The overall architecture, including the FNO block, the encoder-decoder structure, and the integrated attention mechanisms, is illustrated in detail in Figure~\ref{fig3}.

\begin{figure}[H]
\centering
\includegraphics[width=0.9\textwidth]{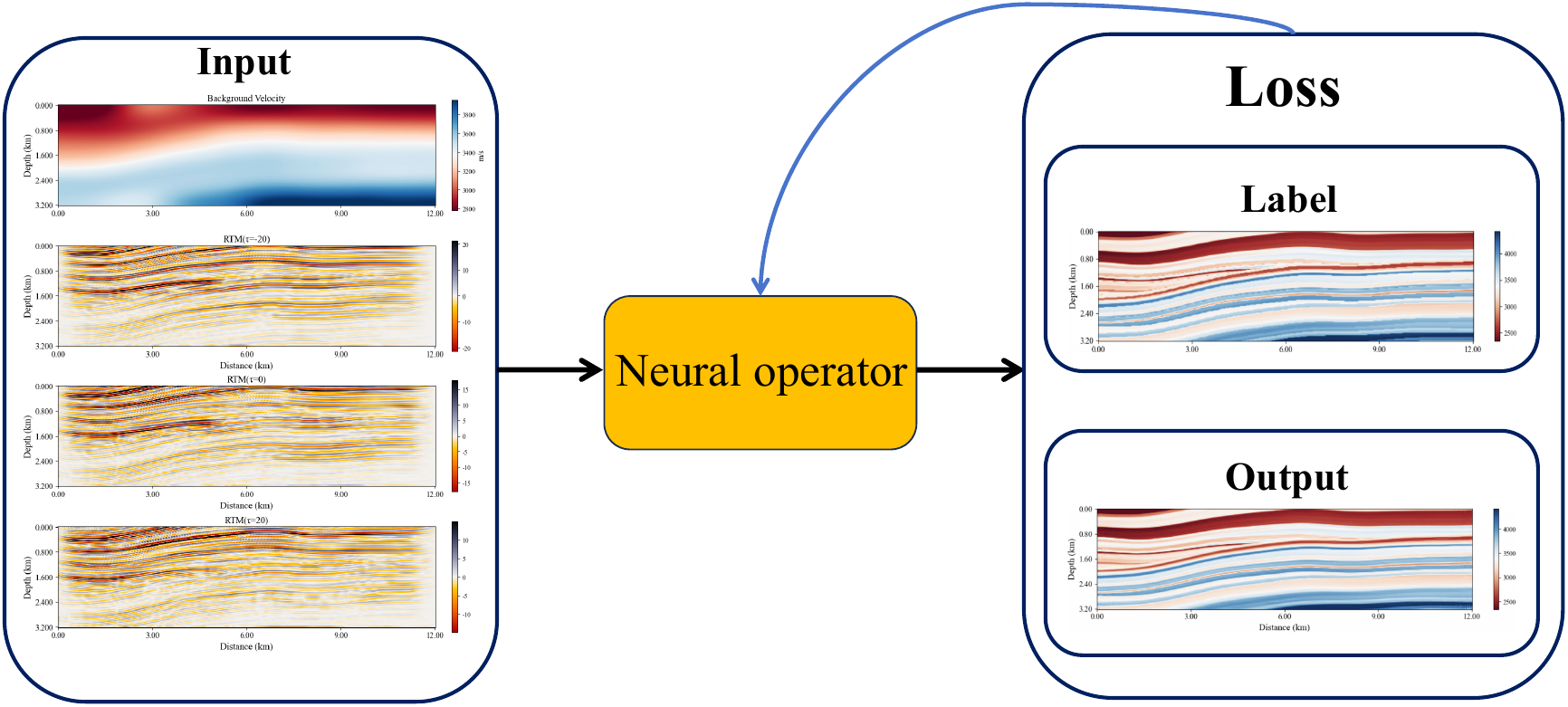}
\caption{\textcolor{black}{During the training stage, the background velocity model together with the corresponding three time-lag images are fed into the neural network. The loss is then computed between the operator’s output and the label, and the parameters of the neural operator are updated through backpropagation until the output closely matches the label.}}
\label{fig3_2} 
\end{figure}

\textcolor{black}{In summary, we use the background (migration) velocity model together with the corresponding time-lag RTM image as the input to the neural operator, while the true velocity model serves as the label. In this way, the network is trained through a supervised learning approach (Figure~\ref{fig3_2}).}

%% file: Sections/Numerical_Examples.tex
\section{Results}
In this section, we first elaborate on the details of the training stage. Following that, in the experimental part, we present a series of case studies using both synthetic and field seismic data to demonstrate the capability of our proposed method in recovering the missing mid-wavenumber components of the velocity models.

\subsection{Training Details}
During the training phase, we employ a total of 6000 synthetic seismic datasets to train and validate the neural operator, with 4800 samples used for the actual training and the remaining for the validation. We use the Adam optimizer to update the network parameters, with a fixed learning rate of $1 \times 10^{-4}$. To balance sensitivity to large errors and robustness to outliers, we utilize a composite loss function that combines the Mean Squared Error (MSE) and the Mean Absolute Error (MAE) as follows:

\begin{equation}
\mathcal{L}_{\text{total}} = 0.9 \, \mathcal{L}_{\text{MSE}} + 0.1 \, \mathcal{L}_{\text{MAE}},
\end{equation}

\begin{figure}
\centering
\includegraphics[width=0.5\textwidth]{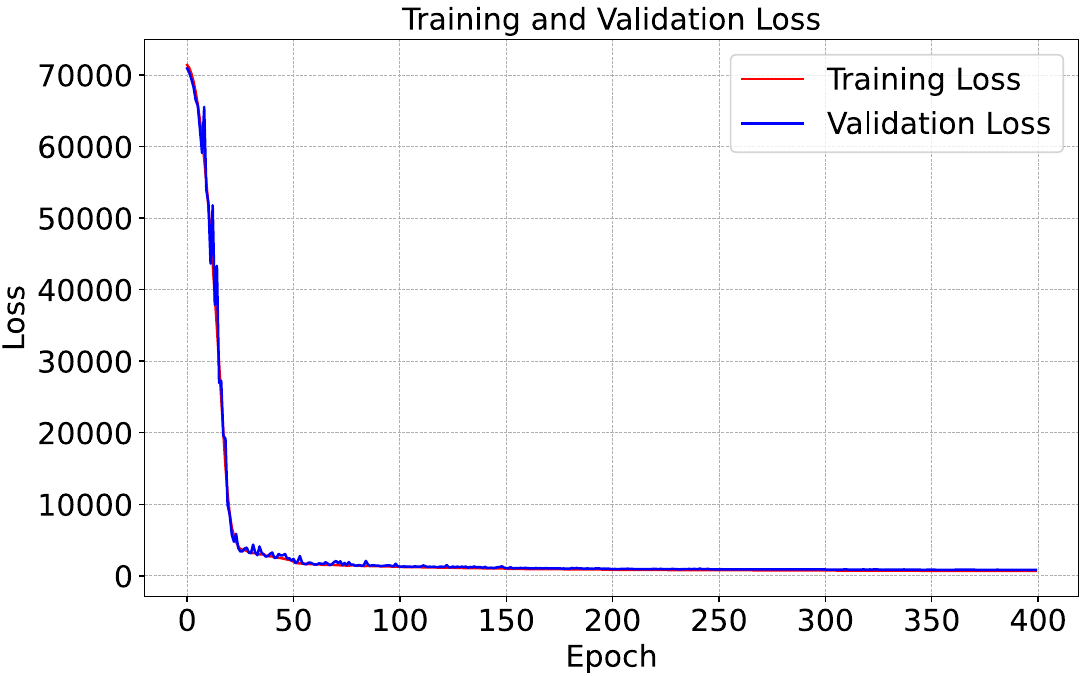}
\caption{Training and validation losses during the training phase.}
\label{fig4} 
\end{figure}

$\mathcal{L}_{\text{total}}$ denotes the total loss used during training, $\mathcal{L}_{\text{MSE}}$ is the mean squared error between the predicted and ground truth data, $\mathcal{L}_{\text{MAE}}$ is the mean absolute error. The weighting coefficients (0.9 and 0.1) reflect the relative importance of MSE and MAE in the final loss. A higher weight is given to MSE to emphasize accurate data fitting, while a smaller weight is assigned to MAE to improve robustness against outliers. The network is trained for a total of 400 epochs, which takes approximately 6 hours to complete. Figure~\ref{fig4} illustrates the evolution of the training loss and validation loss over the course of training. It is clear that both losses decrease steadily and converge to relatively low thresholds, indicating stable training and good generalization performance.

\subsection{In-distribution Test}

\begin{figure}[H]
\centering
\includegraphics[width=\textwidth]{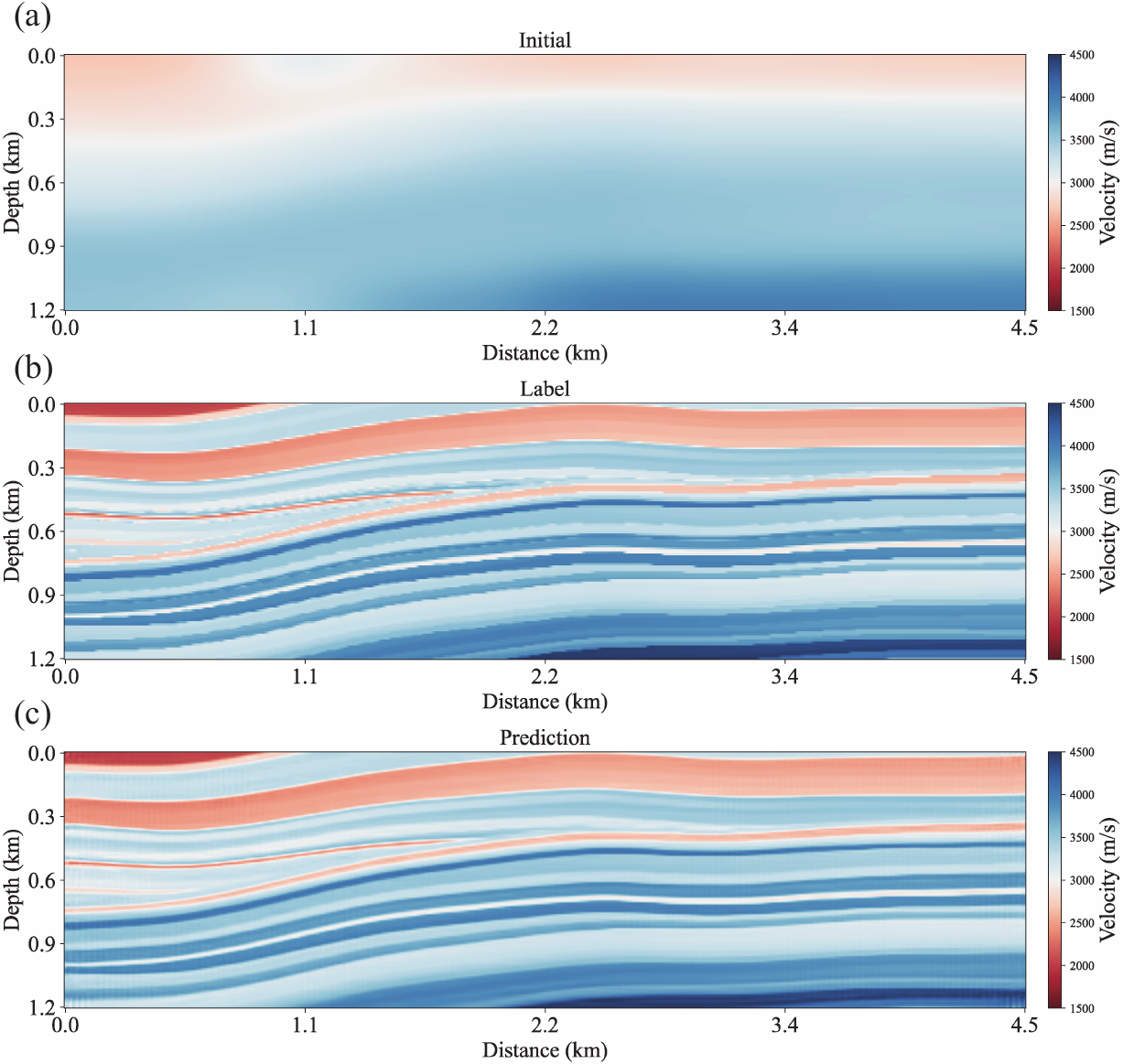}
\caption{An in-distribution test, (a) shows the background (initial) velocity model, displays the corresponding ground-truth velocity label, and (c) presents the predicted result by the neural operator. }
\label{fig5} 
\end{figure}

To evaluate the generalization capability of the trained neural operator, we use in-distribution test data as input to the model. The input consists of an initial velocity model shown in Figure~\ref{fig5}(a) along with its corresponding time-lag RTM images. In comparison with label (Figure~\ref{fig5}(b)) and the neural operator's outputs (Figure~\ref{fig5}(c)), it is evident that the neural operator successfully recovers the missing mid- and high-wavenumber components in the initial velocity model. More specifically, the prediction produced by the neural operator closely matches the ground-truth label across the entire domain. Notably, the model is capable of accurately reconstructing fine-scale stratigraphic features as well as deeper subsurface structures, both of which are typically difficult to resolve. This test confirms that the proposed neural operator exhibits strong mapping capability, accurately reconstructing features from the given input functions.

\subsection{Out-of-distribution Test}

\begin{figure}[H]
\centering
\includegraphics[width=\textwidth]{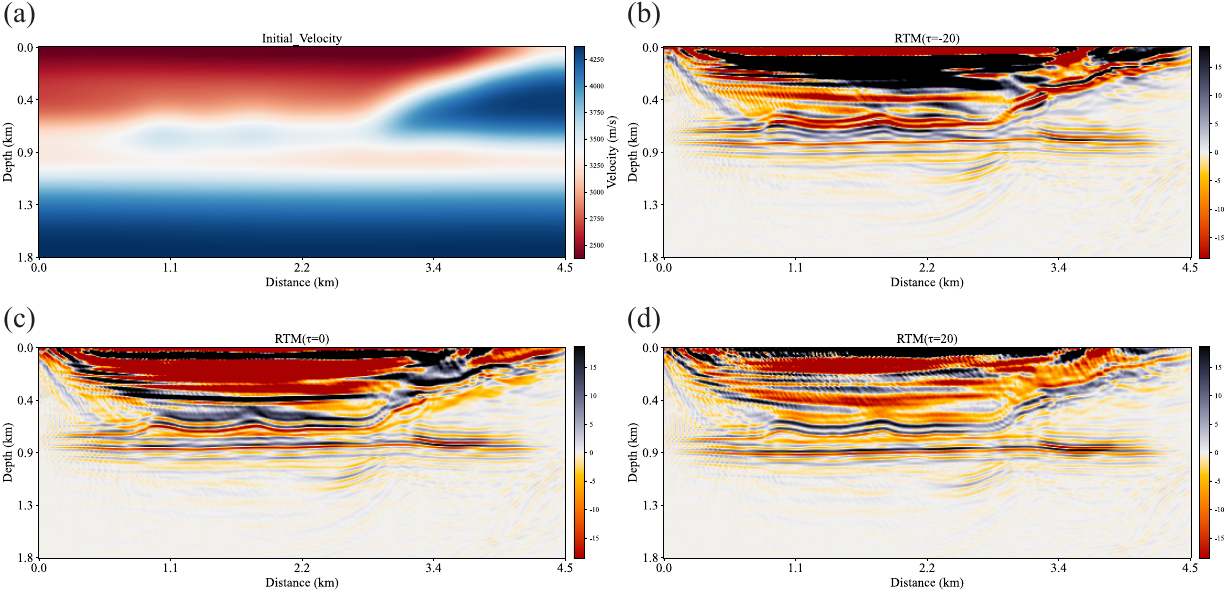}
\caption{To evaluate the neural operator’s capability in predicting high-resolution velocity models under OOD conditions, we input an initial background (migration) velocity model (a) along with three corresponding time-lag RTM images (b-e) to the neural operator. }
\label{fig6} 
\end{figure}

\begin{figure}[H]
\centering
\includegraphics[width=0.8\textwidth]{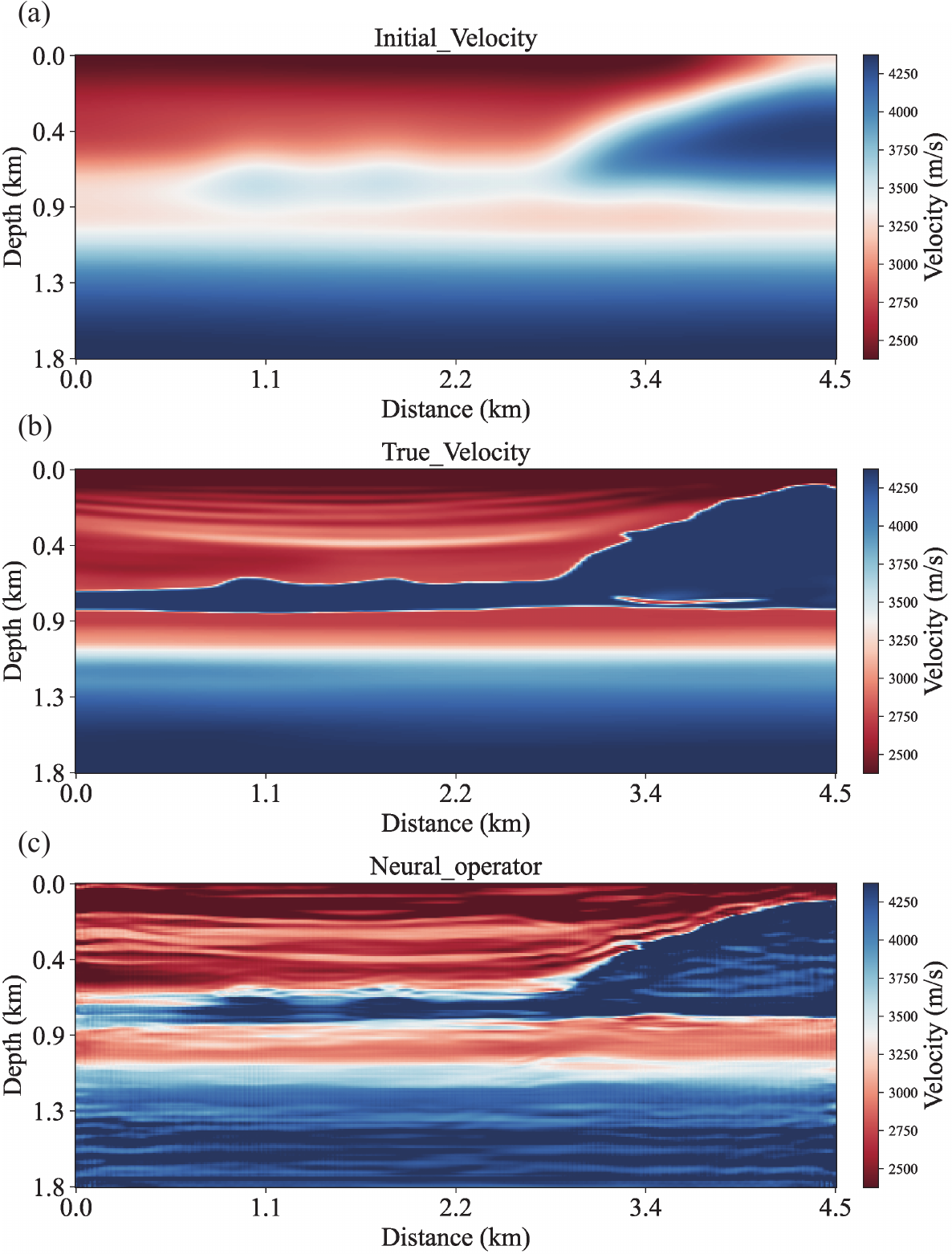}
\caption{An out-of-distribution test, (a) shows the background (initial) velocity model, displays the corresponding ground-truth velocity label, and (c) presents the predicted result by the neural operator. }
\label{fig7} 
\end{figure}

To further evaluate the capability of the neural operator in recovering the missing mid-wavenumber components, we conduct an out-of-distribution (OOD) \textcolor{black}{test using a sample that is developed independently, \textcolor{black}{and thus, lies outside the training distribution (shown in Figure~\ref{fig7}(b)}. Obversely, this sample velocity model differs from those used in the training set with samples shown in Figure~\ref{fig2}.} This data serves as a more challenging scenario to assess the neural operator's generalization ability beyond the seen data. In Figure~\ref{fig6}, we present the four input channels fed into the neural operator. These include an initial velocity model that contains only low-wavenumber information, along with three RTM images corresponding to different time-lag values. Note that in our training dataset, large-scale salt bodies (Figure~\ref{fig7}(a)) that span the entire velocity model are not present, making such structures particularly challenging for our trained model. However, as illustrated in Figure~\ref{fig7}(c), the proposed neural operator successfully reconstructed the overall structure of the salt body, despite it having out-of-distribution features. Moreover, the layered structures beneath the salt body are also accurately recovered, demonstrating the model’s ability to preserve both large-scale and fine-scale features. This example further confirms the effectiveness of the neural operator in recovering high-resolution components, even in geologically complex scenarios. Additionally, we emphasize that the salt body reconstructed by the neural operator exhibits a geologically consistent and structurally complete inversion result. As such, it can serve as a reliable initial model for conventional FWI targeting complex salt structures \citep{alali2023integrating}.

\subsection{Application to Real Seismic Data}

\begin{figure}[H]
\centering
\includegraphics[width=\textwidth]{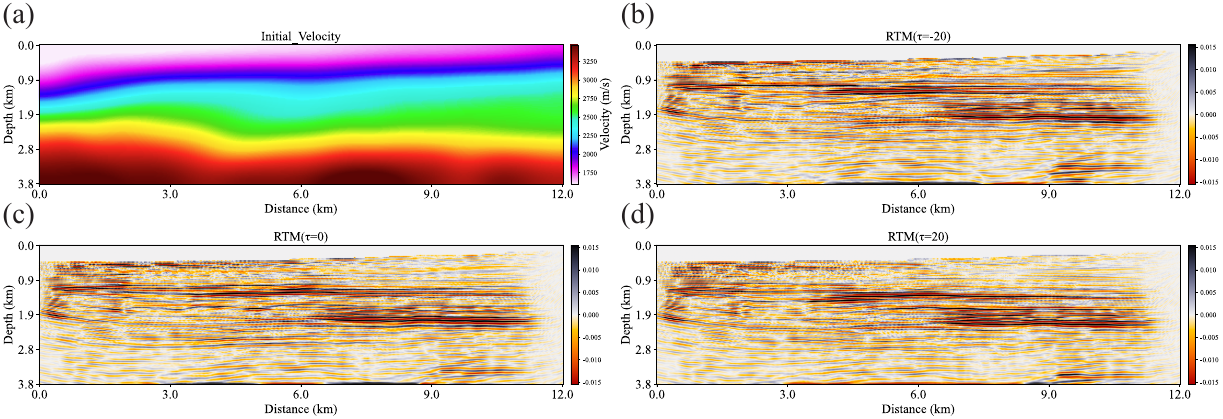}
\caption{The field data test, (a) shows the background velocity model, (b–d) display the corresponding RTM images computed using three different time-lag values.}
\label{fig8} 
\end{figure}

We apply \textcolor{black}{the same fully trained and tested earlier neural operator to a real-world velocity estimate example.} This data is acquired by CGG from the North-Western Australian Continental Shelf. The field seismic data covers a spatial domain of 12 km in length and 3.8 km in depth. A total of 116 shot gathers are evenly distributed along the surface of the velocity model, while 648 receivers are deployed to record the seismic responses. To simulate realistic acquisition conditions and suppress high-frequency noise, a low-pass filter is applied to the seismic wavelet, limiting its frequency content to below 30 Hz. The initial model is obtained using a migration velocity analysis technique, specifically reflection waveform inversion \citep{wu2015full,mu2025full,wang2025implicit}. Again, in Figure~\ref{fig8}, we present the initial (background) velocity model along with the corresponding time-lag RTM images. As shown, by varying the time-lag parameter, the three images capture distinct features of the subsurface reflectivity. Each image emphasizes different aspects of the wavefield, effectively enriching the input space and providing complementary information that is beneficial for velocity model inversion. \par

\begin{figure}
\centering
\includegraphics[width=0.95\textwidth]{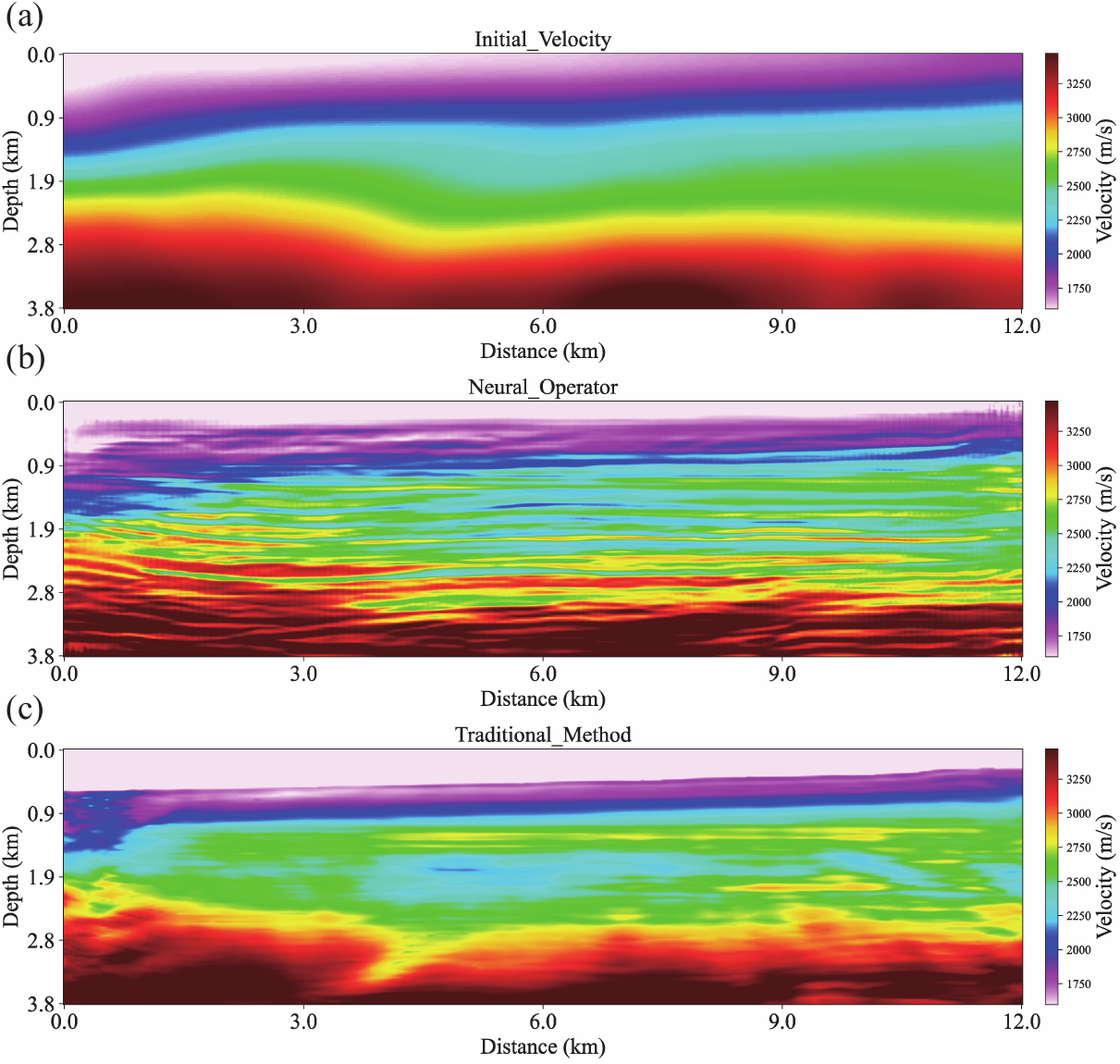}
\caption{(a) shows the background velocity model, (b) presents the high-resolution velocity model predicted by the neural operator, and (c) displays the inversion result obtained using a conventional multi-scale FWI method.}
\label{fig9} 
\end{figure}

Figure~\ref{fig9}(b) displays the inverted velocity model produced by the neural operator. Compared to the initial background velocity model (Figure~\ref{fig9}(a)), the predicted result contains significantly enhanced mid- and high-wavenumber components, leading to improved structural resolution. These high-resolution features are likely attributed to the complementary information provided by the time-lag RTM images as well as the richness of the training dataset. Additionally, Figure~\ref{fig9}(c) displays the velocity model obtained using a conventional multi-scale FWI strategy, employing frequency bands ranging from 3 to 16 Hz. The entire inversion process took approximately 2.5 hours to complete. It is clear that the neural operator produces a prediction result with significantly higher resolution. This comparison clearly highlights the efficiency and accuracy advantages of the proposed neural-operator-based approach over traditional iterative methods. It is also important to highlight that, due to the efficiency of the neural operator’s forward pass, the velocity model was generated in approximately 5 seconds. This demonstrates a substantial computational advantage over conventional FWI, making the approach highly suitable for 3D field velocity building.

\begin{figure}[H]
\centering
\includegraphics[width=0.3\textwidth]{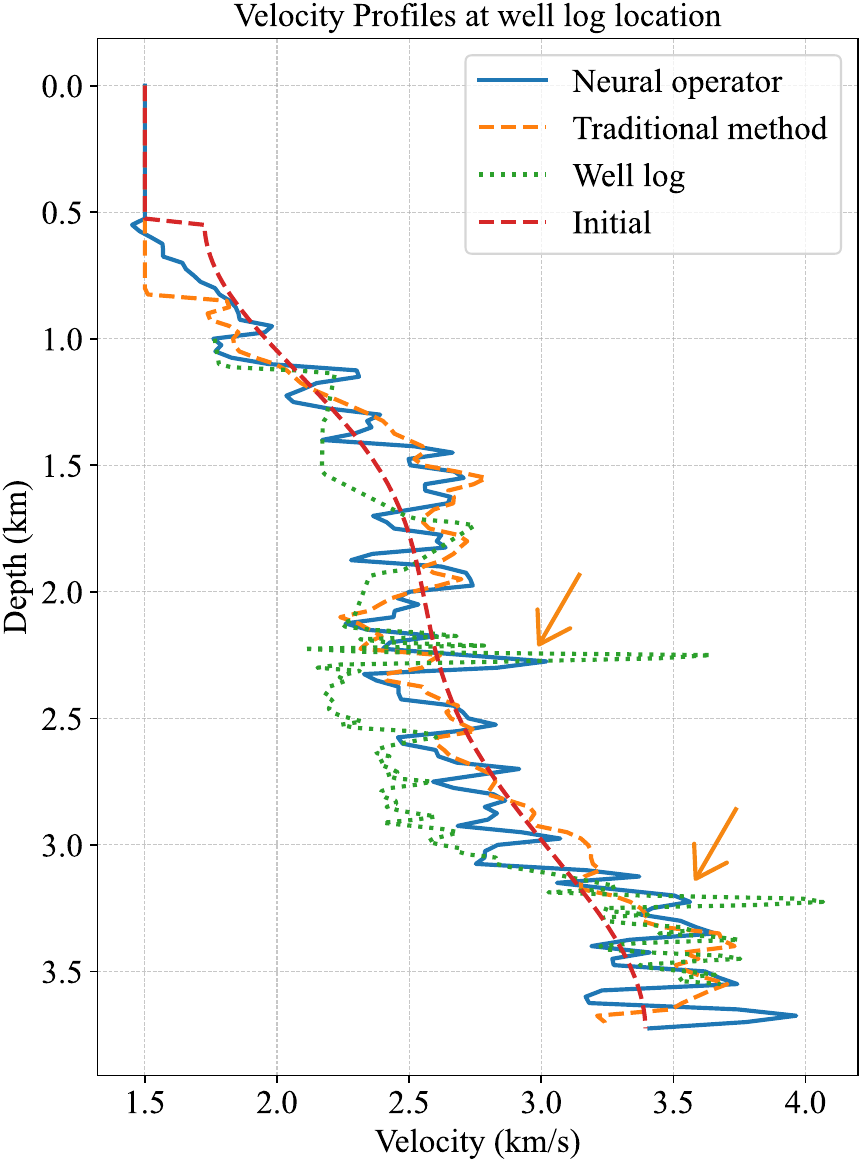}
\caption{ A quantitative well log comparison between the initial background velocity model (red line), the velocity prediction generated by the neural operator (blue line), and the inversion result obtained using a conventional full waveform inversion (FWI) method (yellow line), all benchmarked against the ground-truth well-log data (green line).}
\label{fig10} 
\end{figure}

To further validate the accuracy of the inversion results, we perform a quantitative comparison using well-log data. In Figure~\ref{fig10}, the red curve represents the well-log extracted from the initial velocity model, the blue curve corresponds to the result predicted by the neural operator, and the green dashed line indicates the ground-truth well-log. A close match is observed between the predicted and true velocity profiles, particularly in the regions highlighted by arrows. This agreement demonstrates that the neural operator confirms the reliability of the reconstructed model.

\subsection{Mesh independent test}
\begin{figure}[H]
\centering
\includegraphics[width=\textwidth]{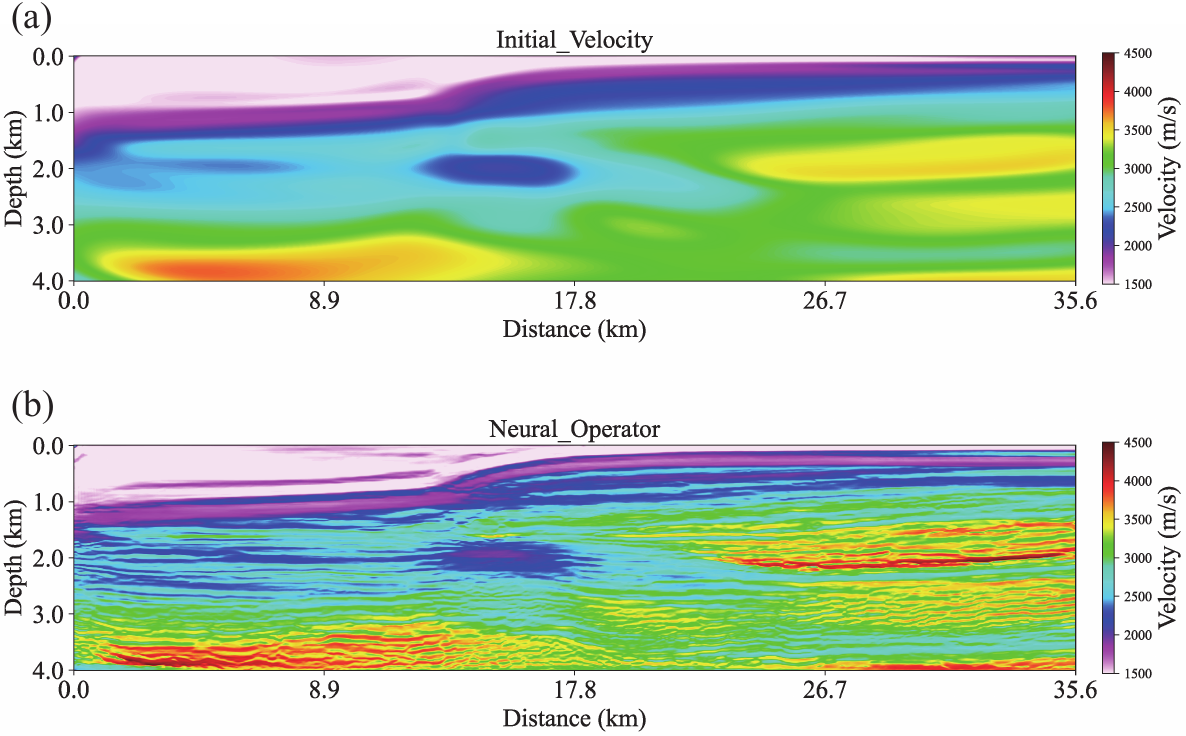}
\caption{A mesh-independent test, we apply the trained neural operator to a larger-scale field dataset without any retraining. (a) presents the background velocity model, while (b) shows the corresponding high-resolution velocity model predicted by the neural operator.}
\label{fig11} 
\end{figure}

\begin{figure}[H]
\centering
\includegraphics[width=0.8\textwidth]{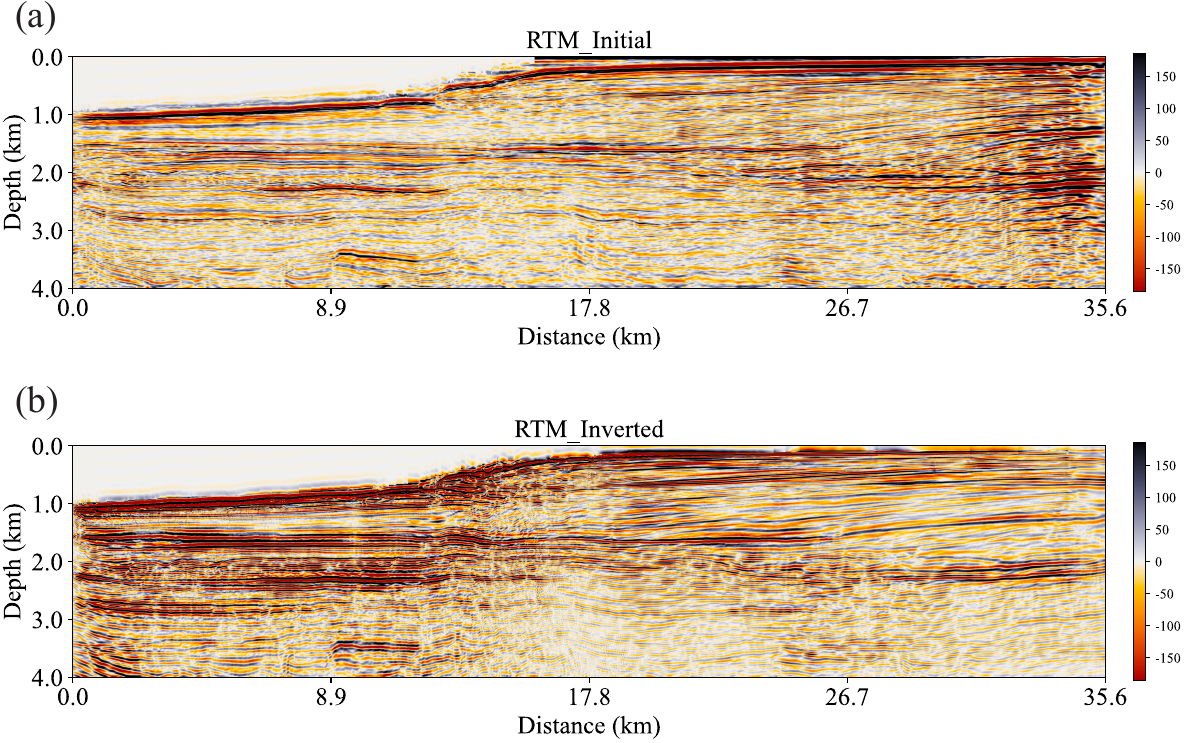}
\caption{(a) shows the RTM image computed using the initial background velocity model, Figure (b) presents the RTM result obtained using the predicted velocity model from our neural operator.}
\label{fig12} 
\end{figure}

The term mesh independence refers to the ability of the neural operator to generalize across different spatial discretizations without requiring retraining. That is, once the model is trained on data defined over a certain grid resolution, it can be directly applied to data defined on finer or coarser meshes. This property arises from the FNO Block in our neural operator, which learns mappings in the function space rather than relying on fixed-dimensional inputs. In particular, we leverage the mesh-independent property of the neural operator to extend its application to the full-scale CGG field data. We directly input the \textcolor{black}{background (migration) velocity model (Figure~\ref{fig11}(a)) and time-lag RTM images} into the neural operator and obtain the inversion result as shown in (Figure~\ref{fig11}(b)). We observe that the neural operator is indeed capable of producing inversion results with considerable structural detail, effectively enhancing the resolution of the reconstructed model. Furthermore, the RTM image obtained by the neural operator's prediction (Figure~\ref{fig12}(b)) exhibits significantly improved structural continuity and enhanced reflector clarity compared to the imaging result using the background velocity model (Figure~\ref{fig12}(a)). 

\section{Discussion}
In this section, we investigate the impact of incorporating time-lag RTM images as additional input channels to the neural operator.
\subsection{The importance of time-lag migration}

\begin{figure}[H]
\centering
\includegraphics[width=\textwidth]{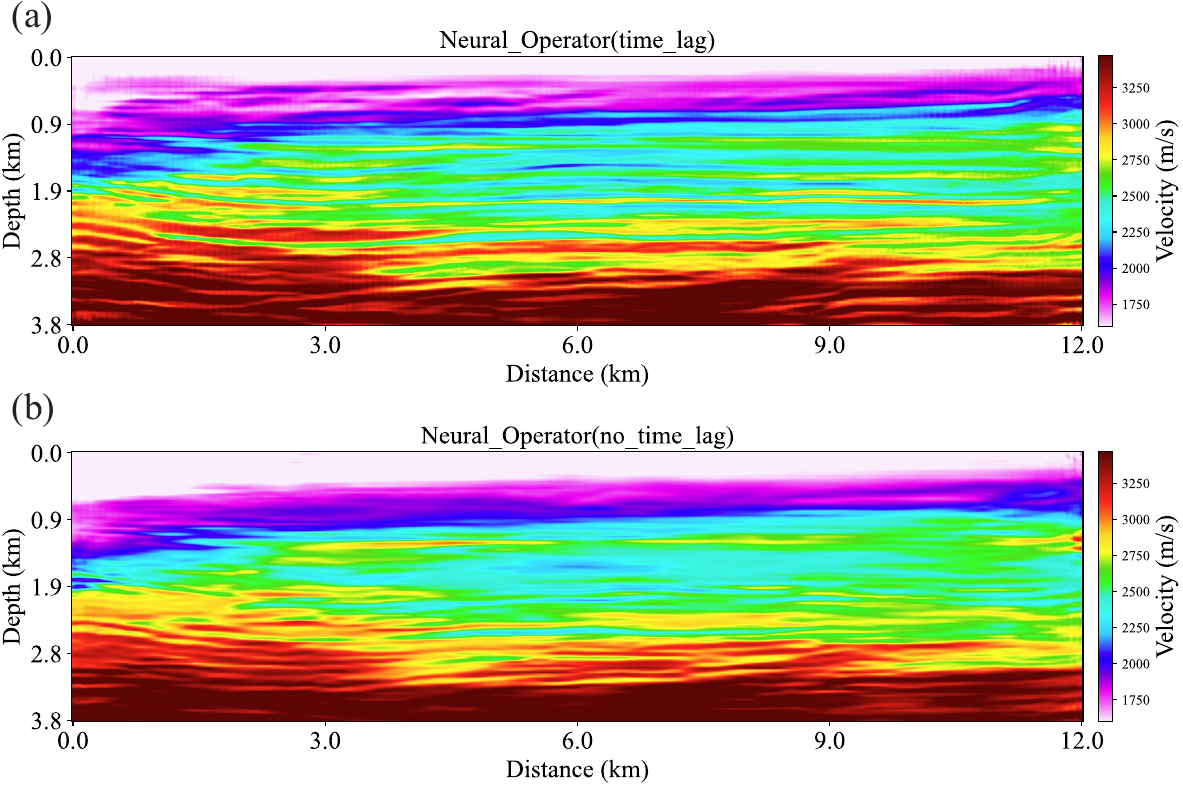}
\caption{(a) and  (b) show the final field data inversion results obtained with and without incorporating time-lag RTM images as input channels, respectively. The comparison indicates that the inclusion of time-lag RTM significantly enhances the resolution and structural accuracy of the predicted velocity model.}
\label{fig13} 
\end{figure}

\textcolor{black}{Time-lag RTM, which is often used for migration velocity model building, serves as a valuable source of additional middle wavenumber information for the neural operator.} As illustrated in Figure~\ref{fig13}, the effectiveness of this input can be clearly observed. Figure~\ref{fig13}(b) shows the inversion result obtained when only two input channels are used—the background velocity and the corresponding RTM image. Compared to the case where time-lag RTM images are incorporated, this result lacks certain structural details, particularly in complex regions. Moreover, the results obtained without incorporating time-lag exhibit lower resolution, \textcolor{black}{but more importantly lower structural consistency,} demonstrating that time-lag RTM inputs provides a certain degree of robustness to inaccuracies in the background velocity.

%% file: Sections/Conclusions.tex
\section{\textbf{Conclusions}}
This study demonstrates the effectiveness of neural operators, particularly the Fourier Neural Operator combined with CNNs, in constructing high-resolution velocity models from time-lag RTM images. By leveraging the mesh-independent properties of neural operators and the capacity to learn mappings in function space, the proposed framework successfully recovers missing mid- and high-wavenumber components in both synthetic and field datasets. The integration of time-lag reverse time migration images as additional input channels significantly enhances the structural resolution and robustness of the inversion results. These extended images provide diverse angle-dependent reflectivity information, which facilitates better learning of complex subsurface features. Experimental evaluations—including in-distribution, out-of-distribution, and mesh-independent tests—demonstrate that the trained neural operator generalizes well under different scenarios. Notably, the model produces prediction results that rival or surpass conventional multi-scale full waveform inversion methods in both resolution and computational efficiency for field data. Overall, this work establishes a promising direction for data-driven seismic inversion by bridging operator learning with advanced image representations. The results highlight the potential of neural operator–based frameworks to serve as efficient, accurate, and scalable alternatives to conventional velocity model build workflows.\\

%% file: Sections/Acknowledgment.tex
\section{\textbf{Acknowledgment}}
This publication is based on work supported by the King Abdullah University of Science and Technology (KAUST). The authors thank the DeepWave sponsors for their support. This work utilized the resources of the Supercomputing Laboratory at King Abdullah University of Science and Technology (KAUST) in Thuwal, Saudi Arabia.
\vspace{0.5cm}

%% file: Sections/CodeAvailability.tex
\section{\textbf{Code Availability}}
The data and accompanying codes that support the findings of this study are available at: 
\url{https://github.com/DeepWave-KAUST/VMB_with_time_lag_RTM}. (During the review process, the repository is private. Once the manuscript is accepted, we will make it public.)